\documentclass[aps,pra,twocolumn]{revtex4-2}
\usepackage{amsmath,epsfig,mathrsfs,graphicx,amssymb,color}
\usepackage[T1]{fontenc}
\usepackage{t1enc}
\usepackage{hyperref}
\usepackage{array}
\usepackage{booktabs}

\def\be#1\ee{\begin{equation}#1\end{equation}}
\newcommand{\ba}{\begin{eqnarray} }
\newcommand{\ea}{\end{eqnarray} }

\usepackage[cp1250]{inputenc}   
\usepackage{amsmath}            
\usepackage{amsfonts}           
\usepackage{amssymb}            
\usepackage{enumerate,graphicx}
\usepackage{array}
\usepackage{booktabs}
\usepackage{yquant}
\usetikzlibrary{fit, quotes}
\usetikzlibrary{snakes,arrows,shapes}
\useyquantlanguage{groups}

\usepackage{physics}

\def\mb{\begin{pmatrix}}
\def\me{\end{pmatrix}}
\def\be#1\ee{\begin{equation}#1\end{equation}}

\begin{document}

\title{Violation of no-signaling  on a public quantum computer}

\author{Tomasz Rybotycki$^{1,2,3}$}
\author{Tomasz Bia{\l}ecki$^{4}$}
\author{Josep Batle$^{5,6}$}
\email{jbv276@uib.es, batlequantum@gmail.com}
\author{Adam Bednorz$^{4}$}

\email{Adam.Bednorz@fuw.edu.pl}

\affiliation{$^1$Systems Research Institute, Polish Academy of Sciences, 6 Newelska Street, PL01-447 Warsaw, Poland}
\affiliation{$^2$Nicolaus Copernicus Astronomical Center, Polish Academy of Sciences, 18 Bartycka
Street, PL00-716 Warsaw, Poland
}
\affiliation{$^3$Center of Excellence in Artificial Intelligence, AGH University,
30 Mickiewicza Lane, PL30-059 Cracow, Poland
}
\affiliation{$^4$Faculty of Physics, University of Warsaw, ul. Pasteura 5, PL02-093 Warsaw, Poland}
\affiliation{$^5$Departament de F\'isica and Institut d'Aplicacions Computacionals de Codi Comunitari (IAC3), Campus UIB, E-07122 Palma de Mallorca, Balearic Islands, Spain}
\affiliation{$^6$CRISP -- Centre de Recerca Independent de sa Pobla, 07420 sa Pobla, Balearic Islands, Spain}

\begin{abstract}
No-signaling is a consequence of the no-communication theorem that states that bipartite systems cannot transfer information
 unless a communication channel exists. It is also a by-product of the assumptions of Bell theorem about quantum nonlocality.
We have tested no-signaling in bipartite systems of qubits from IBM Quantum devices in extremely large statistics,
resulting in significant violations.
Although the time and space scales of IBM Quantum cannot in principle rule out subluminal communications, there is no obvious physical mechanism leading to signaling. The violation  is also at similar level as observed in Bell tests. It is therefore mandatory to check
possible technical imperfections that may cause the violation and to repeat the loophole-free Bell test at much larger statistics,
in order to be ruled out definitively at strict
spacelike conditions.

\end{abstract}

\maketitle

\section{Introduction}

Quantum mechanics violates classical local realism, i.e. a counterfactual definite
local hidden variable model generating measurement results \cite{epr}.
It is  shown by a Bell test \cite{bell}, i.e. violation of a certain inequality,
usually Clasuer-Horne-Shimony-Holt (CHSH) \cite{chsh} or Clauser-Horne \cite{eber}, which requires at least two separated observers, each performing randomly chosen measurements.
The important assumption of local realism is the lack of communication between them, i.e. one party does not know the choice of the other one before accomplishing its own measurement. This assumption cannot be verified \emph{per se}, but its consequences can. The most prominent effect that can be tested is no-signaling, 
that is the result of the measurement
of one party cannot depend on the choice of the other one. Note that it applies to single-party measurements, while two-party correlations 
can depend on both choices, which is the essence of the Bell test. The violation of Bell-type inequalities is a proof of entanglement only when the no-communication
assumption is valid. The other way round, if no-signaling fails, so fails no-communication, and the Bell violation of local realism is meaningless.
Since passing the Bell test is the ultimate proof of entanglement and rejection of local realism, it should be accompanied by a verified no-signaling test.

Experimental Bell tests have a long history of closing detection and communication loopholes \cite{loop1,loop2,larsson}. Detection loophole means that the 
measurement is in fact trichotomic, not dichotomic, common in early optical experiments when the low efficiency of photodetectors lead to high percentage of lost photons, assigned to a third outcome, and causing the whole event to be disregarded. To maintain the Bell conclusion fair sampling  was assumed, i.e.
the counted fraction is representative, not used to invent yet another local hidden variable model \cite{belx1a,belx1b,belx1c,belx1g}. In other implementations, using superconductors, atoms and ions, it is never a problem as the outcome is always dichotomic \cite{belx1d,belx1h,belx1e,belx1f}, although auxiliary photons are sometimes preselected. In contrast to postselection, preselection is fully compatible with the Bell test,
only lowering the overall statistics. Recent Bell experiments, even photonic, have the detection loophole closed \cite{belx2a,belx2b}, but not all \cite{cosm,cosm2,liu}.

The lack of communication can be in principle ruled out by setting the observers, their choices and measurements, within a spatiotemporal framework.
It is commonly assumed that the speed of light is the maximal speed of information transfer, but one has to remember that it does not simply follow
from any, other than free, fundamental relativistic quantum field theory, because it is a nonperturbative claim \cite{b16}. It can be treated as an axiom, consistent with the general expectation of relativistic invariance of fundamental laws \cite{wight,b13}.

\begin{figure}
\includegraphics[scale=.5]{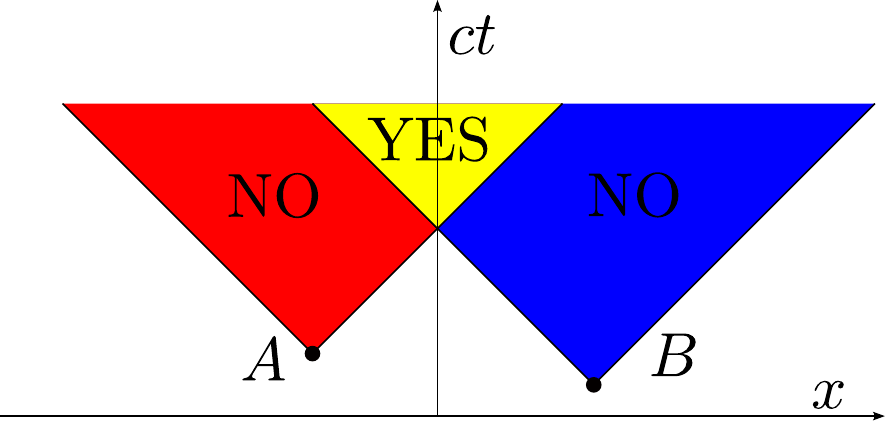}
\caption{Depiction of signaling (YES) and no-signaling regions (NO) in spacetime, here reduced to a single spatial dimension $x$, time $t$ and speed of light $c$.
The choices of $A$ and $B$ marked by black points are the apexes starting the forward causal lightcones (light triangles). Signaling only from $A$ ($B$) is
limited by relativistic causality axioms to the red (blue) region while the yellow part can receive signals from both parties. The remaining white region
neither receives signals from $A$ nor $B$. In the loophole-free Bell test, it is critical to accomplish the measurement within red and blue regions, for $A$ and $B$ 
observers, respectively.}
\label{abs}
\end{figure}

To close the communication loophole, relying on the above axiom, the experimental setup requires sufficient spatial separation between observers so that the accomplished readout  must lie outside of the forward causal lightcone created by the choice of the measurement of the other party, Fig. \ref{abs}.
Although it is compelling from the relativistic point of view, one can still check no-signaling. Certainly, if the axiom is valid, as commonly expected, 
the test should be passed. On the other hand, one can treat no-signaling as a confirmation, of rather lack of falsification, of relativity as regards communication limit. In the recent loophole-free experiments \cite{hensen,nist,vien,munch,storz}, no-signaling is routinely checked. Unfortunately, the present conclusion remains unclear \cite{gisig}. A moderate violation 
of no-signaling occurs  in the tests but has never been checked more accurately \cite{ab17,aden}. A collection of various Bell-type tests \cite{bbt} revealed even
more troubles \cite{sol}. In the first test on superconductors \cite{belx1f}, no-signaling was violated by 70 standard deviations
at extremely large number of trials, $\sim 34\cdot 10^6$, attributed to measurement crosstalk at small distances. The recent loophole-free test \cite{storz} violates no-signaling at the $p$-value (probability that no-signaling hypothesis
holds) of $2\%$ \cite{err} at $\sim 250000$ trials per a pair of choices. Both violations are of the same order so it is tempting to ask what if one reruns the latter test with a much larger number of trials. 

Regarding relativity, it is treated as the ultimate bound on communication, although the physical description of
the loophole-free setups is not directly relativistically covariant (light in the fibers/waveguide travels at about 2/3 of the vacuum speed, 
due to collective interactions in the preferred reference frame)  However, even at small distances and long times, any 
communication needs a reasonable physical origin, an appropriate propagating interaction. In this case violation of no-signaling is helpful in detection
of unspecified communication channels and its analysis can reveal possible interaction mechanism.

Publicly available quantum computers, such as IBM Quantum, offer the real qubits (basic two-level systems, realized on transmons - superconducting Josephson junction shunted with capacitance) \cite{transmon,gam0} and gates (operation on a single qubit or a pair of them, realized by microwave pulses), \cite{gam1,gam2,ecr1,ecr2,ibm,qis}.
Such a computer is expected to realize relatively faithfully the prepared sequences of operations, although they are often noisy, and
cause some crosstalk. Nevertheless, the errors are quite well identified, by thermal noise, leakage to excited states or to the nearest neighbors.
More complicated technical imperfections are expected to be so negligible that can be disregarded. The Bell-type tests can also be run on such computers \cite{alsina,garcia,ibm-bell,ibmb}, although
the communication loophole in the relativistic sense remains open, due to small distances compared to the gate and measurement pulse times. 
Violation of no-signaling is a signature of either serious technical malfunction (e.g. short circuits in cables) or exotic physics behind the scenes.

In this work, we present the results of tests of no-signaling on IBM Quantum devices. They are composed of heavy hexagonal 127-qubit grids
where each qubit is directly connected with one, two, or three other qubits. The connections allow to realize two-qubit gates to create entanglement and
in principle to perform many-qubit operations transpiled into a sequence of native gates.
 We performed three types of experiments, testing signaling between next neighbors and fourth neighbors (parties separated by a chain of 3 other qubits). The nearest, direct, neighbors
 may affect each other by the connection. The experiments are: 
\begin{enumerate}
\renewcommand{\labelenumi}{\alph{enumi})}
\item Bell test on next neighbors, 
\item idle test 
(i.e. local Bell measurements without any entanglement) on next neighbors,
\item idle test on fourth neighbors.
\end{enumerate}
IBM Quantum allows to run tests simultaneously on several pairs of qubits, which is limited by possible path overlapping.
We have found that: i) Bell inequality is violated in the majority of pairs in the test a), ii) no-signaling is violated in all  a-b) tests but it is most prominent
if qubit interlevel frequencies are similar, but still of lower order than Bell violation, iii) violation of no-signaling occurs also in c) but it is much smaller which sometimes requires larger statistics to increase confidence.
In each test, we have found significant violations, at $p$-value below the threshold equivalent to 5 standard deviations, with additional borderline cases, that may become significant if continuing data collection.

We paper is organized as follows. We start with the description of the circuits implemented on IBM Quantum for each test. Next, we present the results,
both Bell and no-signaling tests. Then we explain the commonly suspected origins of signaling which fail to reproduce the observed violations. 
Finally  some conclusions and discussion are drawn. 
Additional technical details are given in Appendix.

\begin{figure}
\begin{tikzpicture}[scale=1]
		\begin{yquant*}[register/separation=3mm]
			init { $\ket 0$} q[0,1,2];
			box {$S$} q[1];
			cnot q[2] | q[1];
			cnot q[0] | q[1];
			cnot q[1] | q[0];
			align -;
			barrier (q);
			box {$S_\alpha$} q[0];
			box {$S_\beta$} q[2];			
			measure q[0];
			measure q[2];
			output {$A$} q[0];
			output {$S$} q[1];
			output {$B$} q[2];
		\end{yquant*}
\end{tikzpicture}
\caption{Standard realization of the CHSH test on IBM Quantum for next neighbors in the test a). The gate $S$ creates a superposition of $|0\rangle$ and $|1\rangle$
states on the source qubits $S$, entangled with the neighbor $B$ by the $CNOT$ gate, and swapped by a pair of $CNOT$s to the other neighbor $A$. The final measurements are the sequences
of $Z_{\alpha/\beta}$ and $S$ gates, $S_\alpha\equiv SZ_\alpha$. In the tests b-c), the entangling part left of the vertical dashed barrier are absent.}
\label{ibmc}
\end{figure}
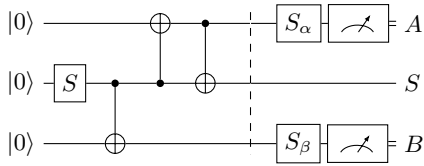

\section{Bell and no-signaling tests}

Implementation of Bell and no-signaling tests on IBM Quantum relies on the grid of qubits, two-level systems with basis states $|0\rangle$ and $|1\rangle$,
in energy eigenspace, differing by the energy $\hbar\omega$,  where $f=\omega/2\pi$ is the drive frequency.
They are manipulated by gates, operations on single or pairs of qubits. 
The  states can be either pure $\rho=|\psi\rangle\langle\psi|$ or mixed, i.e. a convex normalized combination of pure states.
Single qubit states are often represented in the Bloch sphere $\rho=(1+\boldsymbol \sigma \cdot \boldsymbol v)/2$ for the set of Pauli matrices
$\sigma_k$, $k=1,2,3$ and vector $\boldsymbol v=(v_1,v_2,v_3)$ such that $|\boldsymbol v|\leq 1$.

A microwave pulse tuned to the interlevel drive frequency allows one to apply the parametrically controlled gates. The native single qubit gate we use
 is the $\pi/2$ rotation in Bloch sphere about the axis $(1,0,0)$
\begin{equation}
S=\sqrt{X}=(1-i\sigma_1)/\sqrt{2}=\frac{1}{\sqrt{2}}\begin{pmatrix}
		1&-i\\
		-i&1\end{pmatrix}\label{smat},
\end{equation}
in the $|0\rangle$, $|1\rangle$ basis.
The auxiliary $\theta$-rotation about $(0,0,1)$ axis, $Z(\theta)$, is a virtual operation, realized by a phase shift of the next gate, i.e.
\begin{equation}
	S_\theta=Z^\dag_\theta SZ_\theta ,\:
	Z_\theta=
	\begin{pmatrix}
		e^{-i\theta/2}&0\\
		0&e^{i\theta/2}\end{pmatrix}.
\end{equation}
In addition, there is a  two-qubit $CNOT_\downarrow$ gate, operating as
\be
|00\rangle\langle 00|+|01\rangle\langle 01|+|11\rangle\langle 10|+|10\rangle\langle 11|,
\ee
where for $|ab\rangle$ the control qubit states  is $a$ (depicted as $\bullet$) and
target qubit state is $b$ (depicted as $\oplus$ in Fig. \ref{ibmc}). The IBM Quantum devices use
 Echoed Crossed Resonance ($ECR$) gate, instead of $CNOT$ but one can transpile
the latter by additional single-qubits gates, see Appendix \ref{appa}.

For the test a), we create an entangled state, applying $S$ gate to the state $|0\rangle$ of the source qubit, to get
$\sqrt{2}|\psi_0\rangle=|0\rangle-i|1\rangle$ and later $\sqrt{2}CNOT|\psi_0 0\rangle=|00\rangle-i|11\rangle$.
We swap one of qubits to the neighbor by $CNOT_\downarrow CNOT_\uparrow|0\phi\rangle=|\phi 0\rangle$, which holds for an arbitrary $|\phi\rangle$.

The final Bell measurements $A_a$ and $B_b$ are performed by $SZ_\alpha\equiv S_\alpha$ on qubits $A$ and $S_\beta$ on qubit $B$, with the Bell angles
$\alpha=0,\pi/2$ for settings $a=0,1$ and $\beta=-\pi/4,\pi/4$ for settings $b=0,1$, respectively. 
The readout maps projectively the states for the values of observables $A$ or $B$, $|0\rangle\to +1$ and $|1\rangle\to -1$ (we shall abbreviate $\pm 1\to \pm$). 

In the ideal case $\langle A\rangle=\langle B\rangle=0$ while $\langle AB\rangle=-\sin(\alpha+\beta)$, for the average/correlation
defined $\langle x\rangle_{ab}=\sum_x xP_{ab}(x)$ with the $P_{ab}(x)$ being the probability of the outcome $x$ for settings $ab$.

The whole circuit is depicted in Fig. \ref{ibmc}.
We can construct CHSH inequality
\begin{align}
&\bar{CHSH}=\sum_{ab}s_{ab}\langle AB\rangle_{ab}\leq 2,\\
&s_{ab}=\left\{\begin{array}{ll}
+1&\mbox{ for }a=b=0\\
-1&\mbox{ otherwise}\\
\end{array}\right.,
\label{chi}
\end{align}
which is quantum violated at $2\sqrt{2}\simeq 2.818$.

The other tests do not contain the entangling part, just measurements, i.e. operations $S_{\alpha/\beta}$, with the same values $\alpha_{0,1}=0,\pi/2$  and $\beta_{0,1}=-\pi/4,+\pi/4$
and the same or larger distance between $A$ and $B$, as described in Table \ref{tabt}.

\begin{figure*}
\includegraphics[scale=.7]{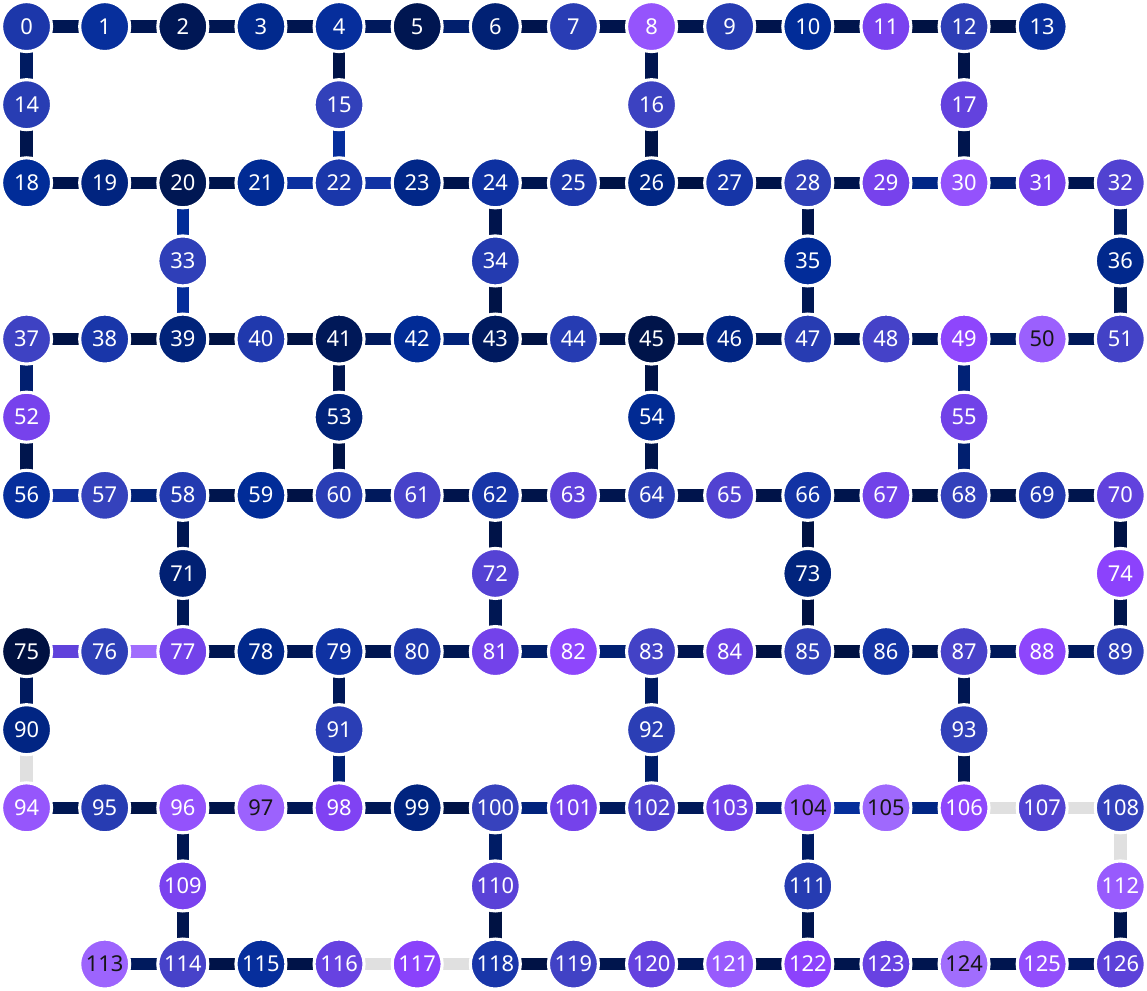}
\caption{Topology of the qubit grid of IBM Quantum devices in Eagle generation, \emph{ibm\_sherbrooke}, \emph{ibm\_brisbane}, \emph{ibm\_kyoto}, \emph{ibm\_kyiv}.
Here the circles represent qubits, bars connections for two-qubit gates. The grid is actually hexagonal.}
\label{top}
\end{figure*}
 
 \begin{table}
	\begin{tabular}{*{3}{c}}
		\toprule
		test& entanglement & $A-B$ distance \\
		\midrule
		a)& yes & 2 \\
		b)& no  & 2 \\
		c)& no  & 4 \\
		\bottomrule
	\end{tabular}
	\caption{Differences between of the tests a-c as regards the entanglement and distance.}
	\label{tabt}
\end{table}

The no-signaling test is performed as follows. In each of a-c) tests, we measure the probability $P_{ab}(AB)$, i.e. how often the pair of values $AB$ are 
measured for a given pair of settings $ab$, and define single-party probability $P(A\ast)=\sum_B P(AB)$, $P(\ast B)=\sum_A P(AB)$. No-signaling holds if
\begin{align}
&\delta P_{a \ast}=P_{a0}(+\ast)-P_{a1}(+\ast),\nonumber\\
&\delta P_{\ast b}=P_{0b}(\ast +)-P_{1b}(\ast +),\label{sig}
\end{align}
are both equal to $0$.
For an ideal implementation $P(A\ast)=P(\ast B)=1/2$ regardless of $AB$ and $ab$. Due to finite statistics, the probabilities are taken from
the actual counts, i.e. $P(x)=N_x/N$, where $N_x$ is the actual number for trials giving the outcome $x$ out of $N$ trials.
It gives a possible error, which can be quantified, assuming
independence of trials. For an equal number of of trials $N$, we have
\begin{align}
&N\sigma^2=N\langle(\delta CHSH)^2\rangle=\sum_{ab}(1-\langle AB\rangle_{ab}^2),\nonumber\\
&N\sigma^2_{a\ast}=N\langle (\delta P_{a\ast})^2\rangle=\sum_b P_{ab}(+\ast)P_{ab}(-\ast)\nonumber\\
&N\sigma^2_{\ast b}=N\langle (\delta P_{\ast b})^2\rangle=\sum_a P_{ab}(\ast +)P_{ab}(\ast -)
\end{align}
where $\delta CHSH=CHSH-\bar{CHSH}$ for $CHSH$ equal the (\ref{chi}) from the actual statistics.
The error is crucial to identify significance of the potential violation of Bell inequality or no-signaling.
In addition, one can express the significance in terms of $p-$value, i.e. the probability that the local realism or no-signaling hypothesis holds.
For a single test it is calculated as the double tail of the Gaussian probability distribution with the above standard deviations, i.e events below  $-|z|$  and above $+|z|$
for the $z$ score, corresponding to the
actually observed value of $CHSH-2$ or $\delta P$, i.e.
\begin{equation}
p(z)=2\int_z^\infty e^{-z^2/2\sigma^2}/\sqrt{2\pi\sigma^2}=\mathrm{erfc}\;(z/\sqrt{2}\sigma).
\end{equation}
The actual $p$-value is taken from the above formula, but multiplied by the number of possible tests, also known as Bonferroni corrections of look-elsewhere
effect \cite{bonf,lee}.

 \begin{table*}
	\begin{tabular}{*{15}{c}}
\toprule
$A-S-B$&a)&CHSH&$\sigma$&$\delta P_{0\ast}$&$\delta P_{1\ast}$&
$\delta P_{\ast 0}$&$\delta P_{\ast 1}$&b)&$\delta P_{0\ast}$&$\delta P_{1\ast}$&
$\delta P_{\ast 0}$&$\delta P_{\ast 1}$&$f_{A-B}$\\
\midrule
55-68-67&&2.25& 3.0&\bf 103&\bf 281&\bf 104&\bf -64.2&&\bf -93.9&\bf -227&\bf -185&\bf -87.6&0.77\\
34-43-44&&2.4& 3.0&\bf 70.8&\bf 63.3&\bf 85&\bf 84.3&&\bf 26.4&\bf -112&\bf -59&\bf 19.1&-4.2\\
29-30-31&&1.84& 3.2&\bf -12.6&6.32&-0.617&\bf -20.3&&\bf -25&\bf 88.2&\bf 89&\bf -23.1&-5.5\\
101-102-103&&2.15& 3.0&\bf -50.1&\bf -76&\bf -63.7&\bf -28.3&&\bf -73.2&\bf 10.2&\bf 62.7&\bf -71.9&6.8\\
7-8-9&&1.6& 3.0&\bf -40.9&\bf -12.7&\bf -13.9&\bf -47.1&&4.55&-3.02&-2.75&3.75&8.2\\
74-89-88&&2.26& 2.8&2&-2.78&-0.305&3.11&&-1.87&\bf -14.2&7.85&\bf 34.4&-9.3\\
94-95-96&&2.14& 3.0&\bf 8.74&6.7&3.98&3.28&&-7.87&-5.25&-4.35&\bf -17.6&15\\
25-26-27&&2.37& 3.0&2.37&-6.63&-1.91&\bf 8.89&&7.83&\bf 10.5&1.53&\bf 16.3&15\\
63-62-72&&2.45& 2.8&0.569&-0.013&-1.11&0.904&&-3.24&2.93&4.9&-4.48&17\\
38-39-40&&2.34& 2.8&0.0717&-7.03&\bf -8.67&\bf -9.78&&-0.827&1.72&0.814&4.08&-17\\
58-59-60&&2.28& 3.0&1.8&-1.53&-1.67&5.53&&\bf 14&2.87&5.38&\bf 18.6&-20\\
21-22-23&&1.59& 3.4&2.92&-1.69&-0.678&3.76&&-0.112&5.26&7.78&-0.222&21\\
80-79-91&&2.19& 3.0&0.352&1.55&4.17&2.34&&1.94&-0.324&0.962&-0.988&-26\\
\bottomrule
\end{tabular}
\caption{Results of the test a) and b) for \emph{ibm\_kyoto}, qubits $A$ and $B$ as specified, with the source qubits $S$ (middle). Here, all $\sigma$ and $\delta P$ are in units $10^{-4}$ while $f_{A-B}=f_A-f_B$ is the frequency difference
between qubit $A$ and $B$ in MHz. The error $\sigma_{a\ast},\sigma_{\ast b}\simeq 1.3\cdot 10^{-4}$.
We have highlighted in bold the strongest violations of no-signaling.}\label{kyo}
\end{table*}

 \begin{table*}
	\begin{tabular}{*{15}{c}}
\toprule
$A-S-B$&a)&CHSH&$\sigma$&$\delta P_{0\ast}$&$\delta P_{1\ast}$&
$\delta P_{\ast 0}$&$\delta P_{\ast 1}$&b)&$\delta P_{0\ast}$&$\delta P_{1\ast}$&
$\delta P_{\ast 0}$&$\delta P_{\ast 1}$&$f_{A-B}$\\
\midrule
1-2-3&&1.68& 3.2&-4.21&0.275&\bf -10.6&2.34&&-1.09&-0.531&-0.24&-0.709&-60\\
7-8-9&&2.15& 2.8&0.072&\bf 21.6&\bf 19.8&4.54&&-1.32&\bf 17.4&\bf 9.61&6.68&-20\\
11-12-13&&2.06& 3.2&\bf 10.4&-2.88&-3.08&0.688&&-0.663&-2.76&-0.184&0.108&-34\\
39-40-41&&2.38& 2.8&-2.44&-1.87&-3.14&-3.63&&\bf 9.75&\bf 9.53&7.08&5.34&-18\\
44-45-46&&2.31& 3.0&-7.11&0.813&\bf 8.66&0.149&&-3.55&0.577&0.951&0.882&96\\
67-68-69&&1.68& 3.2&\bf -55.7&\bf 38.5&\bf 17&\bf -75.9&&\bf -37.6&\bf -14.6&\bf -46&\bf -47.5&-3.7\\
79-80-81&&1.61& 3.2&4.46&-2.55&7.57&-3.06&&-1.62&-0.248&1.64&-2.14&-67\\
83-84-85&&2.39& 3.0&\bf -8.1&-0.238&0.447&-1&&-2.17&1.65&-1.53&1.08&-310\\
95-96-97&&2.18& 3.0&-6.06&-1.76&1.54&-6.21&&-2.48&3.84&6.44&-1.96&23\\
106-107-108&&2.01& 3.3&-0.104&-0.445&-6.61&1.63&&-0.55&2.37&-1.05&-3.26&-180\\
113-114-115&&1.94& 3.3&\bf 78.8&\bf 114&\bf 94&\bf 56.6&&\bf 32.5&71.4&\bf 58.9&\bf 28&4\\
122-123-124&&2.44& 2.8&\bf -10.6&\bf -31.8&\bf -25.2&-2.62&&-6.79&\bf 54&\bf 24.8&\bf -14.6&-12\\
\bottomrule
\end{tabular}
\caption{Results of the test a) and b) for \emph{ibm\_brisbane}, notation as in Table \ref{kyo}}
\label{bri}
\end{table*}

 \begin{table*}
	\begin{tabular}{*{15}{c}}
\toprule
$A-S-B$&a)&CHSH&$\sigma$&$\delta P_{0\ast}$&$\delta P_{1\ast}$&
$\delta P_{\ast 0}$&$\delta P_{\ast 1}$&b)&$\delta P_{0\ast}$&$\delta P_{1\ast}$&
$\delta P_{\ast 0}$&$\delta P_{\ast 1}$&$f_{A-B}$\\
\midrule
4-5-6&&2.06& 3.2&2.66&-0.795&-6.22&2.16&&2.38&1.02&0.362&-2.15&-110\\
11-12-13&&2.25& 3.0&1.6&0.38&1.35&0.881&&-2.08&3.33&-1.39&-1.66&190\\
21-22-23&&1.57& 3.4&\bf 13.9&\bf 14.4&\bf 9.63&\bf 13&&-6.15&4.48&\bf 8.44&-6.7&16\\
28-29-30&&2.23& 3.0&\bf -22.5&6.73&2.41&0.217&&0.156&-0.5&0.324&0.338&51\\
37-38-39&&2.41& 2.8&\bf 8.58&\bf 8.94&3.52&-0.321&&-1.71&\bf -11.1&\bf -10.1&0.926&20\\
43-44-45&&2.54& 2.8&4.65&\bf -82.3&-4.03&\bf 87.2&&\bf 48.1&\bf 56&\bf 59.2&52.7&-7.7\\
47-48-49&&2.6& 2.8&-4.1&\bf -12.7&\bf -18.5&\bf -14.5&&-5.13&\bf -11.6&-7.11&-7.94&-16\\
60-61-62&&2.31& 3.0&0.306&-1.53&-1.2&-0.15&&0.958&-0.0307&-1.05&-1.51&-99\\
80-81-82&&1.24& 3.4&21&\bf -8.86&0.435&-0.775&&-0.627&-0.605&-2.56&1.93&230\\
94-95-96&&2.06& 3.0&1.34&-1.12&1.43&-0.204&&0.021&1.79&1.52&2.29&-130\\
102-103-104&&2.55& 2.8&\bf 133&\bf 198&\bf 101&\bf 29.9&&\bf -12.7&\bf 152&\bf 159&\bf -12.3&2.8\\
117-118-119&&2.53& 2.8&\bf 313&\bf 510&\bf 306&\bf 103&&\bf -295&\bf 249&\bf 242&\bf -308&0.43\\
123-124-125&&2.49& 2.8&-2.02&0.874&-1.38&-1.02&&1.46&2.27&-1.22&0.827&-160\\
\bottomrule
\end{tabular}
\caption{Results of the test a) and b) for \emph{ibm\_sherbrooke}, notation as in Table \ref{kyo}}
\label{she}
\end{table*}
 \begin{table*}
	\begin{tabular}{*{15}{c}}
\toprule
$A-S-B$&a)&CHSH&$\sigma$&$\delta P_{0\ast}$&$\delta P_{1\ast}$&
$\delta P_{\ast 0}$&$\delta P_{\ast 1}$&b)&$\delta P_{0\ast}$&$\delta P_{1\ast}$&
$\delta P_{\ast 0}$&$\delta P_{\ast 1}$&$f_{A-B}$\\
\midrule
11-12-13&&2.36& 4.8&\bf -40.7&\bf 109&0.0432&\bf -145&&\bf -115&\bf -61.9&\bf -70&\bf -109&-1.8\\
80-81-82&&1.82& 5.4&\bf 123&\bf 184&\bf 184&\bf 106&&\bf -17.2&\bf 140&\bf 167&\bf -32.2&-1.8\\
24-25-26&&2.39& 4.8&\bf -166&\bf -85.2&\bf -191&\bf -272&&\bf -91.4&\bf 186&\bf 197&\bf -82.4&-2.6\\
31-32-36&&2.33& 5.0&\bf 22.5&\bf -23.2&\bf 17.1&\bf 65.5&&\bf 14.3&\bf -51.5&\bf -42.8&\bf 12.9&5\\
92-102-101&&2.6& 4.6&\bf -12.7&\bf -66.6&-7.44&\bf 53.3&&3.34&\bf 61.3&\bf 60.6&-0.008&5.2\\
9-8-16&&2.16& 5.0&-4.21&-12.2&-8.86&2.11&&\bf -21.3&\bf 22.6&\bf 28.2&\bf -20.6&5.8\\
38-39-40&&2.21& 5.0&-5.75&\bf -34.9&-1.25&\bf 24.6&&\bf -69.2&\bf 21.8&\bf 19.2&\bf -81.1&-5.8\\
77-78-79&&2.17& 5.0&\bf -30.6&-12.5&\bf -49.3&\bf -71.6&&\bf -42.7&\bf 25.2&\bf 26&\bf -49.7&5.9\\
54-64-63&&2.5& 4.8&\bf -63.1&\bf -17.4&\bf -78&\bf -111&&\bf -58.3&\bf -70.6&\bf -63.8&\bf -49.6&-6.1\\
123-124-125&&2.13& 5.0&\bf 58.7&\bf 64.7&\bf 67&\bf 64.9&&\bf 37.7&2.48&5.58&\bf 43.9&-6.5\\
84-85-86&&1.61& 5.0&6.63&8.89&9.27&4.54&&-10.5&\bf 20.3&\bf 13.5&-6.92&-6.8\\
119-120-121&&2.07& 5.2&\bf 46.9&\bf 36.6&\bf 45.7&\bf 57.2&&\bf -26.3&\bf 31.3&\bf 31.4&\bf -19.1&-9.6\\
15-22-21&&2.41& 4.8&-1.11&2.02&-2.56&-3.84&&-8.84&-3.21&-6.3&\bf -20.5&11\\
\bottomrule
\end{tabular}
\caption{Results of the test a) and b) for \emph{ibm\_kyiv}, notation as in Table \ref{kyo}, except error.
 The error $\sigma_{a\ast},\sigma_{\ast b}\simeq 2.14\cdot 10^{-4}$, and $2.1\cdot 10^{-4}$, in test a) and b) respectively.}
\label{kyi}
\end{table*}
 \begin{table}
	\begin{tabular}{*{6}{c}}
\toprule
$A-B$&$\delta P_{0\ast}$&$\delta P_{1\ast}$&
$\delta P_{\ast 0}$&$\delta P_{\ast 1}$&$f_{A-B}$\\
\midrule
11-31&2.73&0.531&1.03&-1.08&-0.06\\
77-81&-3.43&1.46&-0.557&-1.12&0.073\\
42-59&1.35&-0.644&0.909&-1.69&-1.6\\
116-120&\bf 22.3&\bf 23.3&\bf -22.8&\bf 24.4&3.7\\
68-85&\bf -36.8&-0.803&\bf 19.6&-3.17&8.8\\
15-33&\bf -16.7&\bf -9.9&5.57&-0.193&5.7\\
84-103&-3.09&\bf 9.67&-1.39&\bf 17.6&-6.3\\
97-113&\bf 10&\bf 14.4&\bf 35.7&-5.35&-7.1\\
\bottomrule
\end{tabular}
\caption{Results of the test c) for \emph{ibm\_kyoto}, notation as in Table \ref{kyo}}\label{kyoc}
\end{table}

 \begin{table}
	\begin{tabular}{*{6}{c}}
\toprule
$A-B$&$\delta P_{0\ast}$&$\delta P_{1\ast}$&
$\delta P_{\ast 0}$&$\delta P_{\ast 1}$&$f_{A-B}$\\
\midrule
41-62&-0.425&-0.0373&-0.935&0.216&0.32\\
76-95&2.2&0.705&1.12&1.81&-1.5\\
15-25&-2.18&-0.307&0.982&1.38&2.8\\
7-11&3.52&-2.21&-1.65&-0.457&-4.8\\
28-45&\bf -11.4&\bf -25.5&\bf 20.3&-2.42&-5\\
82-86&-0.231&-0.626&0.86&0.168&-5.2\\
111-125&-0.417&1.7&-4.02&-0.802&-6\\
100-116&\bf -10.6&-1.02&\bf 9.64&\bf -10.9&-7.4\\
\bottomrule
\end{tabular}
\caption{Results of the test c) for \emph{ibm\_brisbane}, notation as in Table \ref{kyo}}
\label{bric}
\end{table}

 \begin{table}
	\begin{tabular}{*{6}{c}}
\toprule
$A-B$&$\delta P_{0\ast}$&$\delta P_{1\ast}$&
$\delta P_{\ast 0}$&$\delta P_{\ast 1}$&$f_{A-B}$\\
\midrule
109-117&-0.284&0.59&-0.0604&-0.775&0.99\\
49-66&-3.36&3.29&1.86&\bf -5.89&1.2\\
68-85&0.102&-0.244&-1.39&0.109&-4\\
92-99&-0.456&0.542&-0.705&-1.7&-4.7\\
34-40&0.293&0.81&0.421&-0.00325&5.2\\
27-46&2.22&-0.283&-0.376&-1.04&6.3\\
120-124&1.53&-1.51&-3.8&1.33&6.5\\
52-71&0.877&-0.484&0.247&0.665&8.6\\
\bottomrule
\end{tabular}
\caption{Results of the test c) for \emph{ibm\_sherbrooke}, notation as in Table \ref{kyo}, except that here  
$\sigma_{a\ast},\sigma_{\ast b}\simeq 6.4\cdot 10^{-5}$}
\label{shec}
\end{table}
 \begin{table}
	\begin{tabular}{*{6}{c}}
\toprule
$A-B$&$\delta P_{0\ast}$&$\delta P_{1\ast}$&
$\delta P_{\ast 0}$&$\delta P_{\ast 1}$&$f_{A-B}$\\
\midrule
97-110&-3.38&-1.41&-0.692&-2.15&0.87\\
26-30&3.42&1.92&-0.00711&0.307&2.2\\
3-14&-0.0507&5.28&-3.34&-1.58&2.6\\
62-66&2.1&-3.9&-3.46&-2.37&-3.6\\
43-60&1.06&-0.298&1.51&-0.499&5.5\\
105-123&0.185&2.29&-3.87&3.44&9.3\\
18-39&\bf -32.3&\bf 26.1&-8.65&1.48&10\\
6-10&2.88&-0.268&-1.15&-0.723&12\\
\bottomrule
\end{tabular}
\caption{Results of the test c) for \emph{ibm\_kyiv}, notation as in Table \ref{kyo}, except that here  
$\sigma_{a\ast},\sigma_{\ast b}\simeq 2.1\cdot 10^{-4}$.}
\label{kyic}
\end{table}

\begin{table}
	\begin{tabular}{*{7}{c}}
\toprule
$ab$&$P(++)$&$P(+-)$&$P(-+)$&$P(--)$&$P(+\ast)$&$P(\ast +)$\\
\midrule
00&0.27779&0.22945&0.26979&0.22297&0.50724&0.54758\\
10&0.28060&0.23202&0.26679&0.22059&0.51262&0.54740\\
01&0.27800&0.22958&0.26975&0.22267&0.50758&0.54775\\
11&0.28092&0.23137&0.26742&0.22029&0.51229&0.54834\\
\bottomrule
\end{tabular}
\caption{Probabilities for $P_{ab}(AB)$ for the test c) on \emph{ibm\_sherbrooke} pair $A-B:49-66$}
\label{prob}
\end{table}

\begin{figure*}
\includegraphics[scale=1]{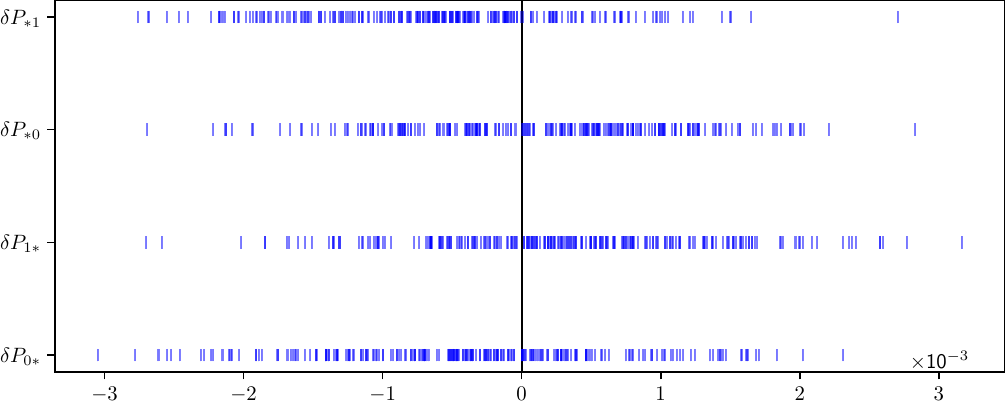}
\caption{Signaling defined by Eq. (\ref{sig}) calculated in the test c) on \emph{ibm\_sherbrooke} pair $A-B:49-66$ for each individual job, i.e.
each bar corresponds to the value calculated for a single job, out of 240.}
\label{scat}
\end{figure*}

\section{Results}

IBM Quantum allows to run experiments in single units, jobs. Each job consists of a sequence of circuits, which can be different.
Each circuit corresponds to an individual experiment run, as specified in the previous section, i.e. a sequence of gates ending with measurements.
The standard time for a single run is 250 microseconds.
Each sequence of circuits is repeated by the number of shots.

We have run tests a-b) on the same sets of qubits, on Eagle generation \emph{ibm\_sherbrooke}, \emph{ibm\_brisbane}, \emph{ibm\_kyoto}, with  20000 shots, 60 jobs,
\emph{ibm\_kyiv} with 7500 shots (due to slower operation), 58 and 60 jobs for  test a) and b) respectively, and
25 repetitions for each choice configuration $ab$, randomly shuffled, giving the total number of circuits 100. We tested simultaneously several non-overlapping pairs of next-neighbor qubits. We have 
depicted the topology of IBM Quantum devices in Fig. \ref{top}. It gives the total number of trials $3\cdot 10^7$ (except $\sim 10^7$ for \emph{ibm\_kyiv}).  The tests c) on the same devices, 
have been run with the same number of jobs (60 for \emph{emph\_kyiv}), shots and repetitions, except \emph{ibm\_sherbrooke}, where the number of jobs was 240.
The results of Bell and no-signaling tests are given in Tables \ref{kyo},\ref{bri},\ref{she},\ref{kyi},\ref{kyoc},\ref{bric},\ref{shec},\ref{kyic}.
The standard deviation in almost all tests is roughly the same, $1.3\cdot 10^{-4}$, except c) on \emph{ibm\_sherbrooke}, $6.4\cdot 10^{-5}$.
The jobs were run in August 2024, expect b) and c) for \emph{ibm\_kyiv} in September 2024, each job takes about $530$ seconds. The total run time was several hours, except c) on \emph{ibm\_sherbrooke},
which was about 3 days. During the test, the devices underwent routine callibrations, which do not affect the experiment, as the test is linear.
In particular, the qubit drive frequencies may vary at the relative level $\sim 10^{-6}$.

It turns out that the majority of tests a) confirm violation of Bell-CHSH inequality, but a-b) also often violate no-signaling.
The violation of no-signaling is the strongest when the frequencies of $A$ and $B$ are similar, but  it still
happens in some cases with large frequency difference, e.g. 80-82 on \emph{ibm\_sherbrooke} and 44-46 on \emph{ibm\_brisbane}
with differences 230 and 96 MHz, respectively.
In all tests c) there are also pairs violating no-signaling, but the violation is smaller, although it seems that still 
larger at small frequency difference.

We have additionally analyzed the extreme case of test c), \emph{ibm\_sherbrooke} pair 49-66, checking the probability, Table \ref{prob},
and results of $\delta P$ for individual jobs, Fig. \ref{scat}. No accidental violation has been found, although the violation
may get some drift over time. The $p-$value is $5.3\cdot 10^{-17}$, taking into account estimating $127\cdot 8$ possible pairs by the look elsewhere effect,
compared to the agreed border at 5 standard deviations, $5.7\cdot 10^{-7}$.
The data and scripts are available publicly \cite{zen}.

\begin{figure}
\includegraphics[scale=.5]{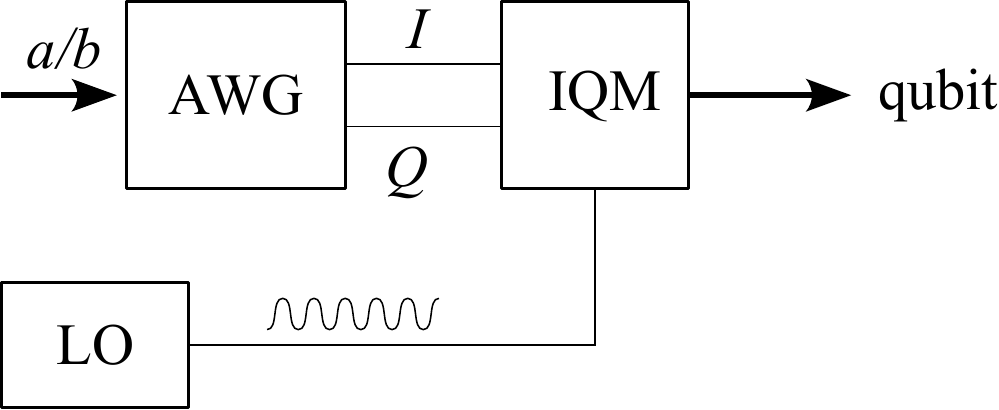}
\caption{The signal flow from the choice $a/b$ to the actual qubits drive pulse.
The AWG creates a pulse amplitude with its in-phase $I$ and quadrature $Q$ component.
The LO creates the continuous wave of the qubit drive frequency. The IQM combines the two waves into a single
microwave pulse fed to the qubit.}
\label{awg}
\end{figure}

\section{Analysis of technical imperfections}

The observed violations of no-signaling are significant, but vary between different qubits pairs. 
It is tempting to seek for the origin in technical imperfections of IBM Quantum devices.
It is known that frequency collisions lead to serious crosstalk, but usually due to two-qubit gates driven by the neighbor's frequency, or
heating disturbance \cite{well,well-sol}.
Certainly the measurement times of $>1$ microsecond compared to the distances of several cm allows communication in the relativistic sense.
However, there are no obvious interactions responsible for it. The most natural $ZZ$ crosstalk, i.e. the interaction diagonal in the energy basis
of the set of qubits, does not help, as $S_{\alpha/\beta}$ differ only by the phase. Even if the qubits have similar frequencies,
they are not exactly equal and the phases are not synchronized. 

If one insists on technical explanations, they must be much more tricky. An error of $S$ gate is insufficient if the $Z$ gate (phase shift)
works correctly. It is the $Z$ gate that must be erroneous, which means an error at the time of pulse preparation.
The microwave pulse is formed by combining Arbitrary Waveform Generator (AWG), Local Oscillator (LO) generating continuous wave of the qubit drive frequency and in-phase/quadrature mixer (IQM),
i.e. the device that combines signals of phase shifted by $\pi/2$ \cite{zgates}, see Fig. \ref{awg}. If the effect is correlation of phase with the pulse amplitude error,
the subsequent $ZZ$ crosstalk may cause signaling. On the other hand, amplitude error would lead to significant local error, i.e.
much more significant change of the qubit directly controlled, not the other party, which we do not observe.

Another brute explanation is a short circuit between cables already before reaching AWG. We refrain from such drastic claims, as most
experiments, even loophole-free Bell tests, rely on the trust in experimental setup. If one cannot take for granted the 
reliable process of the local choice and the time of measurement accomplishements, no conclusions can be drawn from any experiments.
On the other hand,  problems with cables and timings sometimes do become an issue \cite{opera}.

\section{Discussion}

We have checked Bell inequality and no-signaling on IBM Quantum devices.
It turns out that although Bell violation is observed, there is also violation of no-signaling.
The violation of no-signaling is large and cannot explained by a simple crosstalk.
The level of the violation is similar to other superconducting experiments \cite{belx1f,storz}, at very high statistics.
It is urgent to resolve the origin of the violation. Other tests, possibly in different configuration or implementation, should be run.
Also, the loophole-free Bell experiment \cite{storz} should be also rerun at larger number of trials and various configurations 
(also idle, and various sets of angles).
It is certainly difficult to quantify the consequences that these errors may entail in more involved experiments, or how they propagate when global multiqubit tasks are involved. Thus, a thorough further technical analysis to ascertain the exact source of errors is absolutely imperative for future endeavors.
Unless one resolves these issues, more exotic, fundamental explanations
involving extra states beyond simple models predicting extra dimensions, as many worlds/copies \cite{plaga,abadp}, must be considered. 

\section*{Acknowledgments}

The results have been created using IBM Quantum. The views expressed are those of the authors and do not reflect the official policy or position of the IBM Quantum team. We thank Jakub Tworzyd{\l}o for advice, technical support, and discussions, Stanis{\l}aw So{\l}tan for the discussion about frequency collisions, and Witold Bednorz for consultations on error analysis.
TR gratefully acknowledges the funding support by the
program ,,Excellence initiative research university'' for the AGH University in
Krakow as well as the ARTIQ project: UMO-2021/01/2/ST6/00004 and
ARTIQ/0004/2021. We also thank Bart{\l}omiej Zglinicki and Bednorz family for the support.

\appendix

\section{Relation between $CNOT$ and $ECR$ gates}
\label{appa}

\begin{figure}
\begin{tikzpicture}[scale=1]
		\begin{yquantgroup}
			\registers{
			qubit {} q[2];
			}
			\circuit{
			init {$a$} q[0];
			init {$b$} q[1];
			box {$\downarrow$}  (q[0,1]);}
			\equals
			\circuit{
			box {\rotatebox{90}{$CR^+$}}  (q[0,1]);
			box {$X$} q[0];
			box {\rotatebox{90}{$CR^-$}}   (q[0,1]);}
		\end{yquantgroup}
\end{tikzpicture}

\caption{The notation of the ECR gate in the convention $ECR_\downarrow|ab\rangle$}
\label{ecr}
\end{figure}
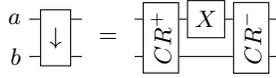
\begin{figure}
\begin{tikzpicture}[scale=1]
		\begin{yquantgroup}
			\registers{
			qubit {} q[2];
			}
			\circuit{
			box {$\uparrow$} (q[0,1]);
			}
			\equals
			\circuit{
			box {$Y_+$}  q[0];
			box {$Y_-$}   q[1];
			box {$\downarrow$}  (q[0,1]);
			box {$H$}  q[0];
			box {$H$}   q[1];
			}
		\end{yquantgroup}
\end{tikzpicture}

\caption{The $ECR_\uparrow$ gate expressed by $ECR_\downarrow$ }
\label{ecrr}
\end{figure}
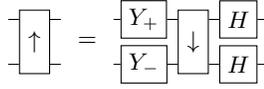

\begin{figure}
\begin{tikzpicture}[scale=1]
		\begin{yquantgroup}
			\registers{
			qubit {} q[2];
			}
			\circuit{
			box {$X$}  q[0];
			box {$S$}   q[1];
			box {$\downarrow$}  (q[0,1]);
			box {$Z_+$}  q[0];
			}
			\equals
			\circuit{
			cnot q[1] | q[0];
			}
		\end{yquantgroup}
\end{tikzpicture}

\caption{The $CNOT_\downarrow$ gate expressed by $ECR_\downarrow$ }
\label{cnot}
\end{figure}
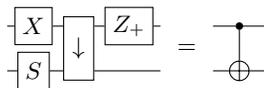
 
 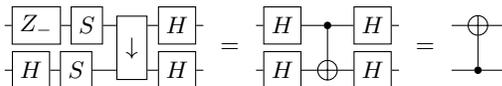
\begin{figure}
\begin{tikzpicture}[scale=1]
		\begin{yquantgroup}
			\registers{
			qubit {} q[2];
			}
			\circuit{
			box {$Z_-$}  q[0];
			box {$H$}   q[1];
			box {$S$}  q[0];
			box {$S$}   q[1];
			box {$\downarrow$}  (q[0,1]);
			box {$H$}  q[0];
			box {$H$}  q[1];
			}
			\equals
			\circuit{
			box {$H$}  q[0];
			box {$H$}   q[1];
			cnot q[1] | q[0];
			box {$H$}  q[0];
			box {$H$}   q[1];
			}
			\equals
			\circuit{
			cnot q[0] | q[1];
			}
		\end{yquantgroup}
\end{tikzpicture}

\caption{The $CNOT_\uparrow$ gate expressed by $ECR_\downarrow$ }
\label{cnotr}
\end{figure}

The IBM Quantum devices use a native two-qubit $ECR$ instead of $CNOT$ \cite{ecr1,ecr2} but one can transpile the latter by the former, adding single qubits gates.
We shall use Pauli matrices in the basis $|0\rangle$, $|1\rangle$,
\be
X=\begin{pmatrix}
0&1\\
1&0\end{pmatrix},\:Y=\begin{pmatrix}
0&-i\\
i&0\end{pmatrix},\:Z=\begin{pmatrix}
1&0\\
0&-1\end{pmatrix},\:
I=\begin{pmatrix}
1&0\\
0&1\end{pmatrix}.\label{pauli}
\ee
We also denote two-qubits gates by $\downarrow$ and $\uparrow$, which mean the direction of the gate (it is not symmetric), i.e.
$\langle a'b'|G_\uparrow|ab\rangle=\langle b'a'|G_\downarrow|ba\rangle$.

The $ECR$ gate acts on the states $|ab\rangle$ as (Fig. \ref{ecr})
\ba
&ECR_\downarrow=(XI-YX)/\sqrt{2}=CR^- (XI) CR^+=\nonumber\\
&
\begin{pmatrix}
0&X_-\\
X_+&0\end{pmatrix}
=\begin{pmatrix}
0&0&1&i\\
0&0&i&1\\
1&-i&0&0\\
-i&1&0&0\end{pmatrix}/\sqrt{2},
\ea
in the basis $|00\rangle$, $|01\rangle$, $|10\rangle$, $|11\rangle$
where the native gate is
\be
S=X_+=X_{\pi/2}=(I-iX)/\sqrt{2}=\begin{pmatrix}
1&-i\\
-i&1\end{pmatrix}/\sqrt{2},
\ee
and $X_-=X_{-\pi/2}=ZX_+Z$, 
with 
\be
CR^\pm=(ZX)_{\pm \pi/4},
\ee
using the convention  $V_\theta=\exp(-i\theta V/2)=\cos(\theta/2)-iV\sin(\theta/2)$ if $V^2=I$ or $II$.
The gate is its inverse, i.e. $ECR_\downarrow ECR_\downarrow=II$.

Note that $Z_\theta=\exp(-i\theta Z/2)=\mathrm{diag}(e^{-i\theta/2},e^{i\theta/2})$ is a virtual gate adding essentially the phase  shift to next gates.
 $ECR$ gates can be reversed, i.e., for $a\leftrightarrow b$, (Fig. \ref{ecrr})
\be
ECR_\uparrow=(IX-XY)/\sqrt{2}=(HH) ECR_\downarrow(Y_+Y_-),
\ee
denoting $V_\pm= V_{\pm \pi/2}$, and Hadamard gate,
\be
H=(Z+X)/\sqrt{2}
=Z_+SZ_+=\begin{pmatrix}
1&1\\
1&-1\end{pmatrix}/\sqrt{2},
\ee
and $Z_\pm SZ_\mp=Y_\pm$, with $Y_+=HZ$ and $Y_-=ZH$.

The $CNOT$ gate can be expressed by $ECR$ (Fig. \ref{cnot})
\ba
&CNOT_\downarrow=(II+ZI+IX-ZX)/2=\nonumber\\
&
\begin{pmatrix}
I&0\\
0&X\end{pmatrix}
=\begin{pmatrix}
1&0&0&0\\
0&1&0&0\\
0&0&0&1\\
0&0&1&0\end{pmatrix}\nonumber\\
&
=(Z_+ I)ECR_\downarrow (XS),
\ea
while its reverse reads (Fig. \ref{cnotr})
\ba
&CNOT_\uparrow=(II+IZ+XI-XZ)/2
=\nonumber\\
&\begin{pmatrix}
1&0&0&0\\
0&0&0&1\\
0&0&1&0\\
0&1&0&0\end{pmatrix}=(HH)CNOT_\downarrow(HH)\nonumber\\
&
=
(HH)ECR_\downarrow (SS)(Z_-H).
\ea


\begin{thebibliography}{99}
\bibitem{err}
There is a typo in \cite{storz} in Table SVI, missing 0 after period in rows 2 and 4.
\end{thebibliography}
\end{document}